# Mobility Transition at Grain Boundaries in Two-Step Sintered 8 mol% Yttria Stabilized Zirconia

Yanhao Dong and I-Wei Chen*

Department of Materials Science and Engineering, University of Pennsylvania, Philadelphia, PA 19104, USA

**Abstract**

Stagnation of grain growth is often attributed to impurity segregation. Yttria-stabilized cubic zirconia does not evidence any segregation-induced slowdown, as its grain growth obeys the parabolic law when the grain size increases by more than one order of magnitude. However, lowering the temperature below 1300 °C triggers an abrupt slowdown, constraining the average grains to grow by less than 0.5 μm in 1000 h despite a relatively large driving force imparted in the fine grains of ~0.5 μm. Yet isolated pockets of abnormally large grains, along with pockets of abnormally small grains, emerge in the same latter sample. Such microstructure bifurcation has never been observed before, and can only be explained by an inhomogeneous distribution of immobile four-grain junctions. The implications of these findings for two-step sintering are discussed.

**\*Corresponding Author Information**

**Tel:** +1-215-898-5163; **Fax:** +1-215-573-2128

**E-mail address:** iweichen@seas.upenn.edu (I-Wei Chen)

**Postal address:** Department of Materials Science and Engineering, University of Pennsylvania, LRSM Building, Room 424, 3231 Walnut St., Philadelphia, PA 19104-6272

I.   Introduction

Two-step sintering has been practiced in a variety of ceramics, including $Y_2O_3$[1,2], $Al_2O_3$[3-5], doped $ZrO_2$[6,7], $ZnO$[8], $TiO_2$[9], $BaTiO_3$[10-12] and Ni-Cu-Zn ferrite[10]. The idea is rooted in the hypothesis that grain growth can be suppressed in a sinterable system without necessarily suppressing grain boundary diffusion.[1,2] This is because grain growth in a polycrystal requires the motion of the entire grain boundary network, which includes 2-grain boundaries, 3-grain lines and 4-grain junctions. If the sintering temperature is so chosen that the 3-grain lines and 4-grain junctions are immobile, yet grain boundary diffusion is still active to allow pore densification and two-grain boundary flexing, then it is possible to achieve densification without grain growth. This is schematically illustrated in **Fig. 1** using 4-grain junction as an example. In practice, one uses the first step of a higher temperature $T_1$ to shrink the pores to a size relatively small compared to the grain size; this places the pores and the ceramic in a thermodynamically sinterable state, typically ~70-75% relative density. It is followed by the second step of lowering the temperature to $T_2$ to achieve full densification without grain growth. While two-step sintering is a rather unconventional modification to the standard lab sintering procedure, it is actually quite compatible to the industrial practice, which subjects ceramic bodies to a slow heating/cooling cycle of 10 to 48 hours passing through various parts of a tunnel furnace. Since two-step sintering has produced some of the finest and most uniform ceramic microstructures[1-12], and along with it, improved reliability[13], it is of great technological interest. A more detailed understanding of its kinetics is thus merited, which motivated this study.

Clearly, **Fig. 1** is critical for two-step sintering. Yet there is so far no data to support it in relevant ceramic systems. Therefore, the aim of this study is to obtain such data in yttria stabilized zirconia (YSZ) via grain growth studies. Cubic YSZ is selected as the model system for the following reasons. First, unlike its tetragonal counterpart, cubic YSZ has minimum solute drag and experiences rather fast grain growth[14-16]. For example, with the common sintering practice of 1400-1500 °C for a few hours, a grain size of ~5 μm is easily obtained as opposed to a grain size of only ~0.3 μm in the case of tetragonal YSZ. Therefore, if two-step sintering can successfully produce a fine-grain cubic YSZ, it would not only reaffirm the utility of the approach but also provide a set of samples uncommonly suitable for grain growth study. This is because the smaller the starting grain size, the larger the driving force and the wider the temperature window to determine grain boundary/junction mobilities, especially at lower temperatures when mobilities are low. Second, YSZ has a fluorite structure, which is known to be defect-tolerant.[17] As a result, many cations can dissolve in YSZ in very large amounts making the growth kinetics relatively insensitive to impurities. Third, there is relatively little crystallographic anisotropy in cubic YSZ, making it unlikely to exhibit grain boundary faceting that may profoundly influence grain-growth thermodynamics and kinetics. Indeed, no abnormal grain growth, which is often correlated to grain faceting and solute segregation, has ever been reported in cubic YSZ. Finally, parabolic grain growth predicted by the standard grain growth theory of Hillert[18] has been confirmed many times in dense fluorite-structure oxides and their derivatives, including YSZ[14,16], $CeO_2$[19,20] and $Y_2O_3$[21], which portends well for our grain growth study in cubic YSZ; it also provides a data base to compare the grain boundary and junction mobilities in different ceramics of this family.

To set the stage of the kinetic study, below we briefly outline the kinetic competition between 2-grain boundaries, 3-grain lines and 4-grain junctions. (Detailed solutions of the following models are available elsewhere.[22]) If the growth of grain size $G$ is limited by the mobility $M_b$ of 2-grain boundaries, then Hillert's growth velocity[18] driven by a capillary pressure $2\gamma/G$ with $\gamma$ as the interfacial energy applies

$$\frac{dG}{dt} = 2M_b\gamma\left(\frac{1}{G_{cr}} - \frac{1}{G}\right) \quad (1)$$

Here, $G_{cr}$ is the critical grain size that neither grows nor shrinks at the time $t$, which sets up a chemical potential $2\gamma/G_{cr}$ to ensure mass (volume) conservation. Extending the above to the case when the grain velocity is limited by the mobility of 3-grain line, $M_t$, we may write

$$\frac{dG}{dt} = 2M_t\gamma\left(\frac{G}{a}\right)\left(\frac{1}{G_{cr}} - \frac{1}{G}\right) \quad (2)$$

Here, we assume the driving force on a grain boundary of an area $G^2$ is entirely spent on a 3-grain line, which has an effective area of $aG$ with $a$ taken as the atomic spacing. Likewise, if the grain velocity is limited by the mobility of 4-grain junction, $M_j$, then

$$\frac{dG}{dt} = 2M_j\gamma\left(\frac{G}{a}\right)^2\left(\frac{1}{G_{cr}} - \frac{1}{G}\right) \quad (3)$$

Here, we assume the entire driving force is spent on a 4-grain junction with an effective area of $a^2$. In mixed control, setting all the above velocity the same and letting them share the total driving force $2\gamma(1/G_{cr}-1/G)$, one can show that

$$\frac{dG}{dt} = 2\gamma \left( \frac{1}{M_b} + \frac{a}{M_t G} + \frac{a^2}{M_j G^2} \right)^{-1} \left( \frac{1}{G_{cr}} - \frac{1}{G} \right) \qquad (4)$$

We next estimate the scaling laws for growth in the context of mean-field-theory, taking $(1/G_{cr} - 1/G)$ to be of the order of $1/G$. It follows that $G^2 \sim M_b \gamma t$ or parabolic growth holds for 2-grain boundary control, $Ga \sim M_t \gamma t$ or linear growth holds for 3-grain line control, and $\ln(G/G_0) \sim M_j \gamma t/a^2$ (where $G_0$ is a reference grain size) or exponential growth holds for 4-grain junction control. (Parabolic law has been verified many times in the past in dense ceramics[16,19-21], some evidence of exponential growth was also seen during post-sintering annealing of two-step-sintered $Y_2O_3$ of 100 nm grain size[2].) Lastly, while in mixed control the growth is expected to evolve through several transient stages, eventually parabolic growth will dominate since the concentrations of 3-grain lines and 4-grain junctions will rapidly decrease with the grain size. In the above, the transient times may be estimated from the above scaling laws by setting $G$ to the prevailing grain size $G^*$.

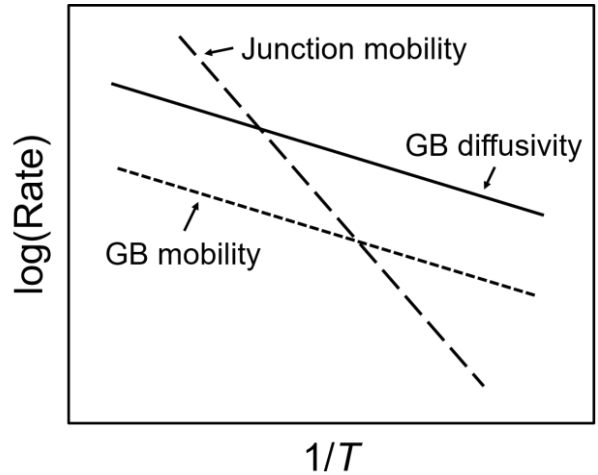

**Figure 1** Schematic Arrhenius plot of grain boundary (GB) diffusivity, GB mobility and junction mobility. GB diffusivity and GB mobility are assumed to have similar activation energy, while junction mobility has higher activation energy.

## II. Experimental procedures

Pressed pellets of 8YSZ powders (TZ-8Y, Tosoh Co., Tokyo, Japan) were isostatically compacted at ambient temperature. For normal sintering (NS), the 300 MPa compacted pellets were heated to 1300 to 1500 °C at 5 °C/min, then held for various time (**Table 1**). For two-step sintering (TSS), the 1 GPa pressed compacts were heated to $T_1$=1290 °C at 10 °C/min, then immediately cooled down to $T_2$=1200 °C and held for 16 h. Sintered pellets were fully dense and uniform (**Fig. 2a-c**). However, to minimize data scattering, only cut pieces from two "parent" dense pellets were used for subsequent grain growth experiments in air: A 1300 °C NS sample with a starting grain size $G_0$ of 1.7 μm (**Fig. 2a**) for 1300-1450 °C growth study, and a TSS sample with a $G_0$ of 0.49 μm (**Fig. 2c**) for 1175-1300 °C growth study. After experiments, samples were polished and thermally etched at 1220 °C for 0.2 h to reveal grain boundaries, examined under a scanning electron microscope (SEM, Quanta 600, FEI Co. Hillsboro, OR). The linear intercept method with over 100 intercepts and a correction factor of 1.56 was used to measure the average grain size $G_{avg}$. To obtain the grain size distribution, we first manually outlined the grain boundaries, then calculated the mean radius (the average radial distance from the centroid to the boundary, measured at 2° interval) for 200-1100 grains using Image-Pro Plus (Media Cybernetics, Inc., Rockville, MD).

**Table 1 Required sintering conditions and obtained grain sizes of dense 8YSZ by normal sintering (NS) and two-step sintering (TSS)**

| Sintering method | Sintering condition | Grain size (μm) |
| --- | --- | --- |
| NS | 1500 °C for 2 h | 5.8±0.2 |
| | 1450 °C for 2.5 h | 4.6±0.3 |
| | 1400 °C for 4 h | 3.2±0.2 |
| | 1350 °C for 6 h | 2.6±0.3 |
| | 1300 °C for 12 h | 1.7±0.05 |
| TSS | $T_1$=1290 °C for 0 h, $T_2$=1200 °C for 16 h | 0.49±0.03 |

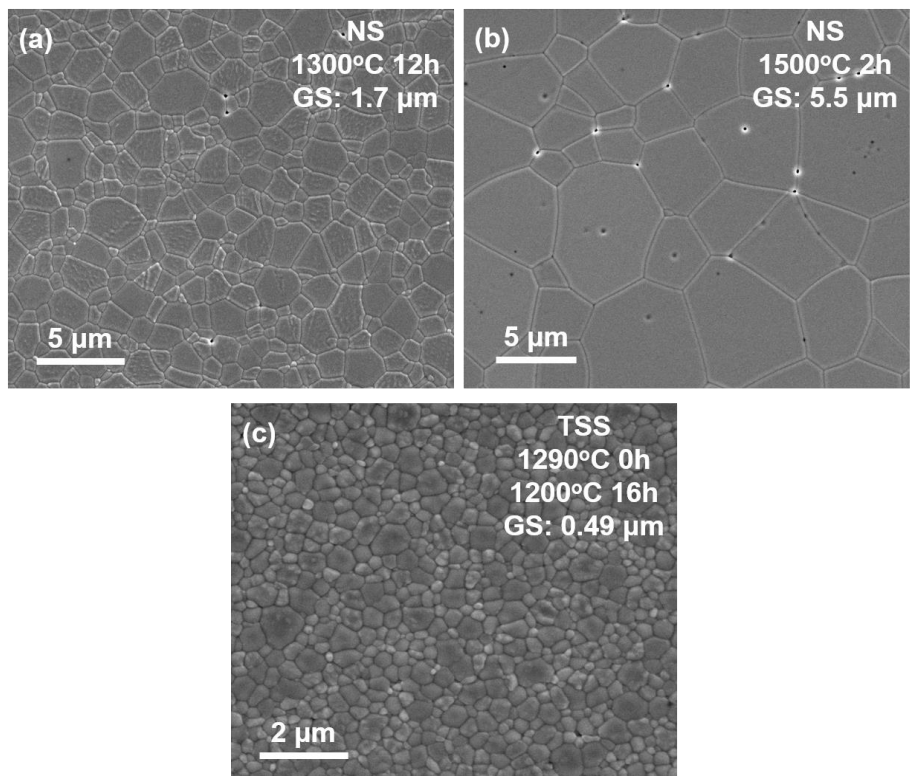

**Figure 2** Microstructures of as-sintered 8YSZ by normal sintering (NS) at (a) 1300 °C for 12 h and (b) 1500 °C for 2 h, and by two-step sintering (TSS) at (c) $T_1$=1290 °C for 0 h followed by $T_2$=1200 °C for 16 h. Grain sizes listed in upper right corners.

### III. Results

(1) Grain Growth

The NS sample with $G_0$ of 1.7 μm follows the parabolic growth law with time $t$, **Fig. 3**,

$$G_{\text{avg}}^2 - G_0^2 = 2M\gamma t \qquad (5)$$

Here $\gamma$ is the grain boundary energy, taken as 0.3 J/m$^2$ in this work. The microstructures after annealing remain uniform, two examples of which are shown in **Fig. 4** after 4 h's growth at 1300 °C and 1450 °C, respectively. The calculated grain boundary mobility $M$ (red triangles in **Fig. 5**) follows the Arrhenius relationship with an apparent activation energy $E_a$ of 4.2 eV, which is typical for the family of fluorite structure oxides[14,19,21,24]. At 1300 °C, the TSS sample with $G_0$ of 0.49 μm also follows the parabolic law with the same slope over a growth time of 64 h in **Fig. 3**, growing $G_{\text{avg}}$ to 5.7 μm. This datum point is shown in **Fig. 5** as a blue triangle, which coincides with the other data at 1300 °C (red triangle), confirming the same mobility in both samples despite their different "parents". These mobility data in **Fig. 5** are in line with the literature data of 8YSZ[14], which may be compared with those of other fluorite structured oxides (summarized in **Fig. 12** of Ref. 23).

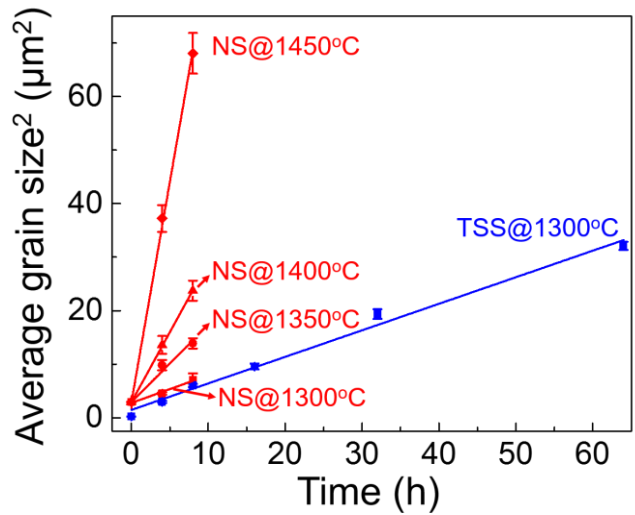

**Figure 3** Parabolic grain growth of average grain size in 8YSZ. Initial grain size: 1.7 μm (red, NS) and 0.49 μm (blue, TSS).

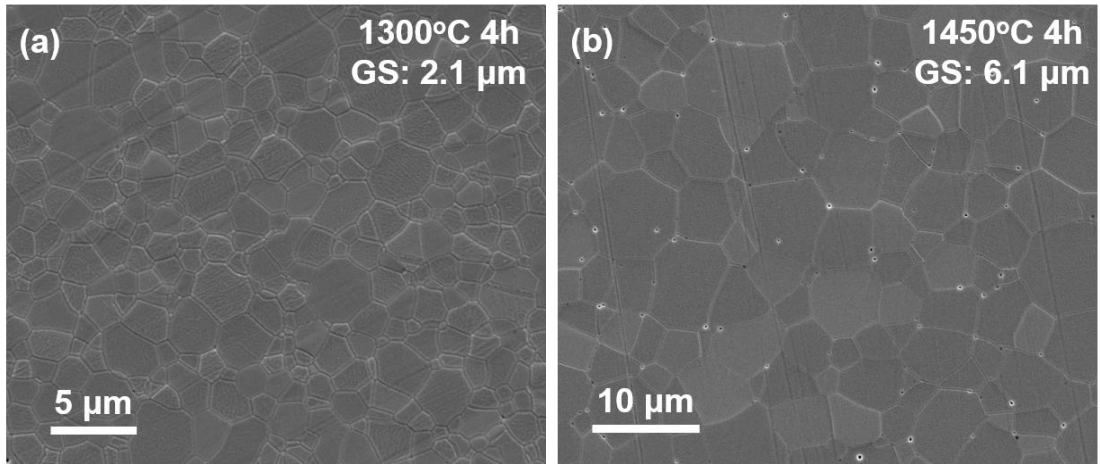

**Figure 4** Microstructures of 8YSZ NS samples after post-sintering annealing at (a) 1300 °C for 4 h and (b) 1450 °C for 4 h. Initial microstructure shown in **Fig. 2a**.

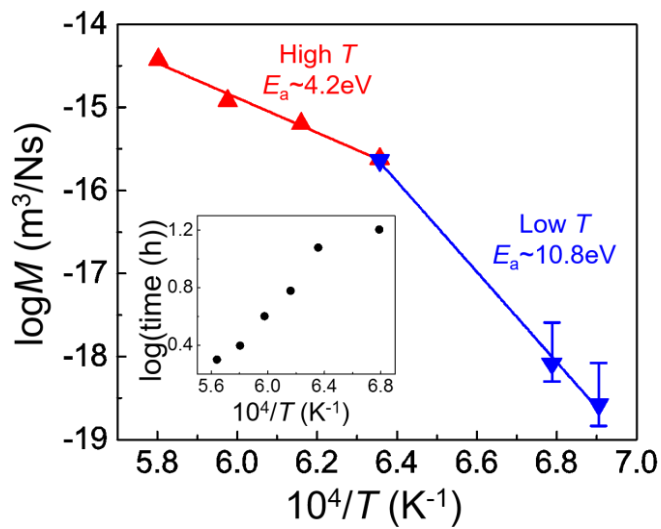

**Figure 5** Arrhenius plot of grain boundary mobility of 8YSZ. Red: NS-parent sample, blue: TSS-parent sample. Inset: Sintering time-temperature data from **Table 1**.

Grain growth below 1300 °C was studied using the TSS sample, whose $G_0$=0.49 μm provides a larger driving force. Nevertheless, $G_{avg}$ only reached 0.94 μm after 1000 h at 1175 °C (**Fig. 6a**) and 1.07 μm after 500 h at 1200 °C (**Fig. 6b**). On the $G_{avg}^2$ vs. $t$ plot in **Fig. 7**, stagnation departing from the parabolic growth is evident, the more so the lower the temperature.

This is also evident from a power-law fit to the data using

$$G_{\text{avg}}^n - G_0^n = At \qquad (6)$$

which gives $n=4.9$ at 1200 °C and $n=6.7$ at 1175 °C.

Despite the slowdown, we have force-fitted $G_{\text{avg}}^2$ vs. $t$ using dashed lines in **Fig. 7** to define an upper-bound mobility from the short-time data and a lower-bound mobility from the long-time data. These data are plotted in **Fig. 5** where the error bars are specified by the two bounds. Together with the mobility at 1300 °C, they suggest an extremely high $E_a$ of 10.8 eV. Phenomenologically, it signals a rapidly quenched average mobility below 1300 °C, which is reminiscent of **Fig. 1**, as it divides the temperature into two regimes, the lower one being the window for two-step sintering. The feasibility of low-temperature sintering is consistent with the sintering-time data of **Table 1**, displayed in the inset of **Fig. 5** showing no obvious slowdown.

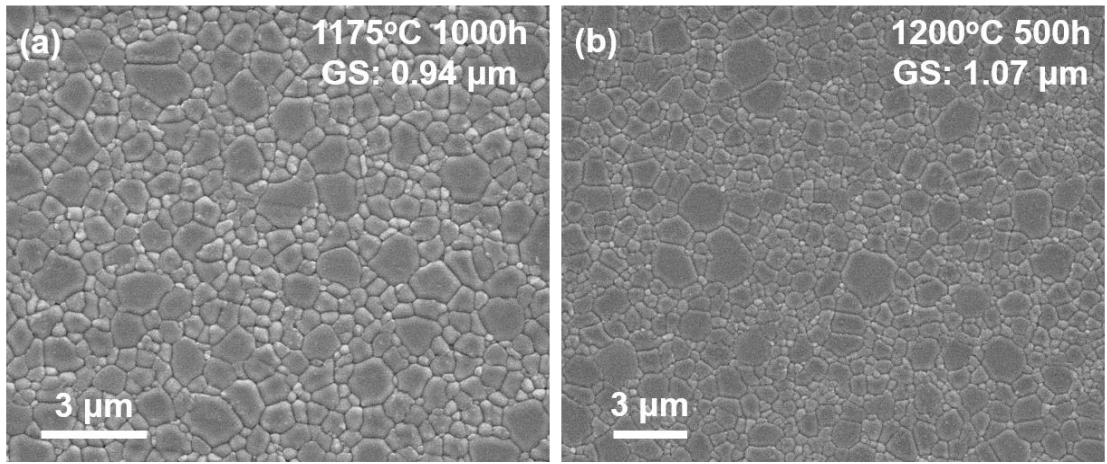

**Figure 6** Microstructures of 8YSZ TSS samples after post-sintering annealing at (a) 1175 °C for 1000 h and (b) 1200 °C for 500 h. Initial microstructure shown in **Fig. 2c**.

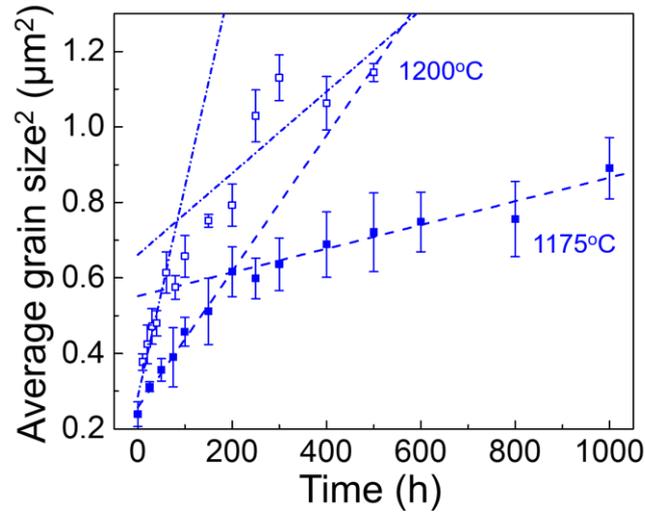

**Figure 7** Grain growth data represented as (average grain size)$^2$ vs. post-sintering annealing time at 1175 °C and 1200 °C. Dashed lines are linear fittings for first/last six datum points. Initial grain size: 0.49 μm (TSS).

(2) Microstructural inhomogeneities

While the microstructure of two-step sintered samples initially remained uniform during post-sintering annealing, some inhomogeneities began to emerge after prolonged low-temperature annealing as shown in **Fig. 8**. Starting with the microstructure in **Fig. 2c** and $G_0$=0.49 μm, this sample was annealed at 1175 °C for 500 h to grow to $G_{avg}$=0.85 μm with a relatively uniform microstructure shown in **Fig. 8a**. However, within the same sample of a matrix $G_{avg}$=0.85 μm, there is also a nano-grain cluster of many <200 nm grains, shown in in **Fig. 8b**. Elsewhere in the same sample of again a matrix $G_{avg}$=0.85 μm, in **Fig. 8c**, there is another cluster, this time of large grains, some reaching 4.1-5.1 μm, i.e., 5-6 times the $G_{avg}$. Similar inhomogeneities were found in other long-time-annealed samples (micrographs not shown), and they only emerged after at least 500 h at 1175 °C or at least 200 h at 1200 °C. Therefore, microstructural inhomogeneities did develop at low temperatures after a long incubation time despite very modest overall growth.

In **Fig. 9d** these abnormalities at 500 h are quantified by their grain size distributions set in the background of the prevailing grain size distribution in the rest of the sample. They are clearly "outliers" compared to the majority grains, centered around $G_{avg}$, at $\log(G/G_{avg})=0$. It is also interesting to note that the as-sintered TSS sample (**Fig. 9c**) has a narrower grain size distribution, as given by the standard deviation $\sigma$, than either the 500 h annealed sample (**Fig. 9d**) or the 1000 h annealed (**Fig. 9e**). On the other hand, the NS samples developed progressively narrower distributions as grain growth proceeds; for example, the NS 1400°C-8 h-annealed sample (**Fig. 9a**) has a narrower distribution than that of the NS 1300°C-8 h-annealed sample (**Fig. 9b**). The breadth of these distributions can also be compared in terms of the maximum grain size (i.e., the upper cut-off): In TSS samples it starts from $2.5G_{avg}$ ($\log(G/G_{avg})=0.4$) before annealing and grows to $4G_{avg}$ ($\log(G/G_{avg})=0.6$) after long-time annealing at 1175 °C (and 1200 °C, data not shown); in contrast, it never exceeds $2.5G_{avg}$ ($\log(G/G_{avg})=0.4$) in NS samples despite much more grain growth starting from $G_0=1.7$ μm.

The diverging trends toward inhomogeneities in terms of the standard deviation $\sigma$ of the normalized grain size $G/G_{avg}$ are plotted in **Fig. 10** against the dimensionless time, which, according to Eq. (1), may be defined as $\tilde{t} = 2M_b\gamma t = (G_{avg}/G_0)^2 - 1$. Above 1300 °C, the standard deviation gradually decreases with $\tilde{t}$ starting from the as-sintered NS sample, but below 1300 °C it actually increases slightly with $\tilde{t}$ starting from the as-sintered TSS sample. Since $\sigma$ is a statistical measure of microstructural inhomogeneity, it is clear that two-step sintering offers the most uniform microstructure ($\sigma=0.43$) among as-sintered ceramics, but it degrades during extended, lower-temperature post-sintering annealing accompanied by non-parabolic growth. In contrast, when annealing is conducted at a higher temperature, parabolic grain growth will sharpen the grain size distribution to eventually reach $\sigma<0.45$.

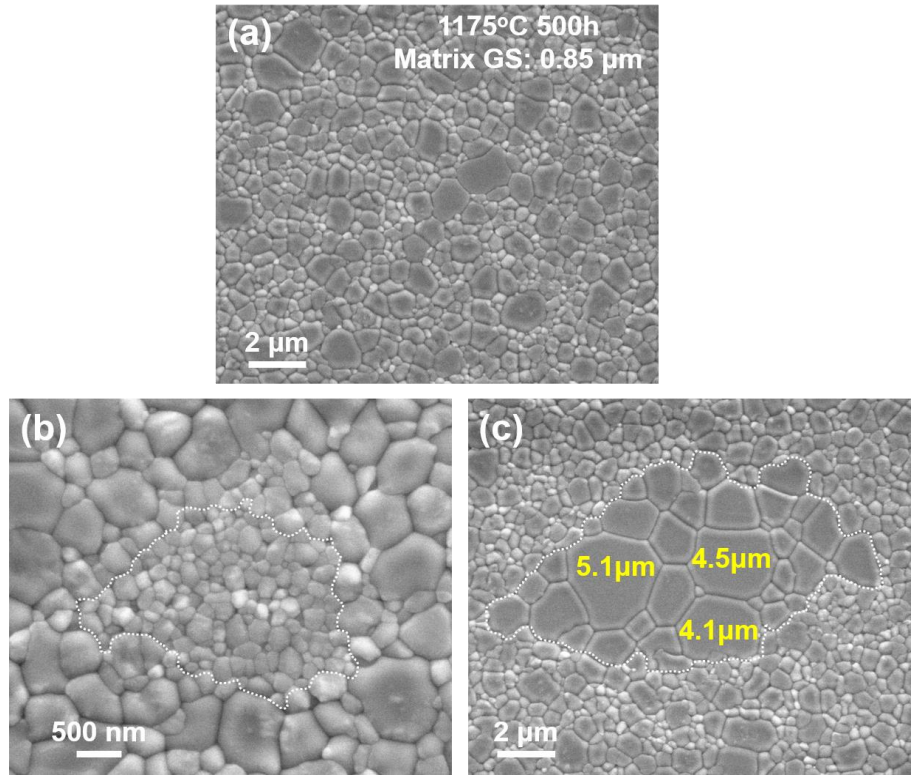

**Figure 8** Three microstructures developed in 1175 ºC-500 h annealing in same TSS sample, $G_0$=0.49 μm (**Fig. 2c**). (a) Normal region with uniform microstructures; $G_{avg}$=0.85 μm. (b) Nano-grain cluster of <200 nm grains; matrix grain size=0.85 μm. (c) Large-grain cluster; matrix grain size=0.85 μm. Grain size distributions of circled regions in (b-c) shown in **Fig. 9c** as dotted curves.

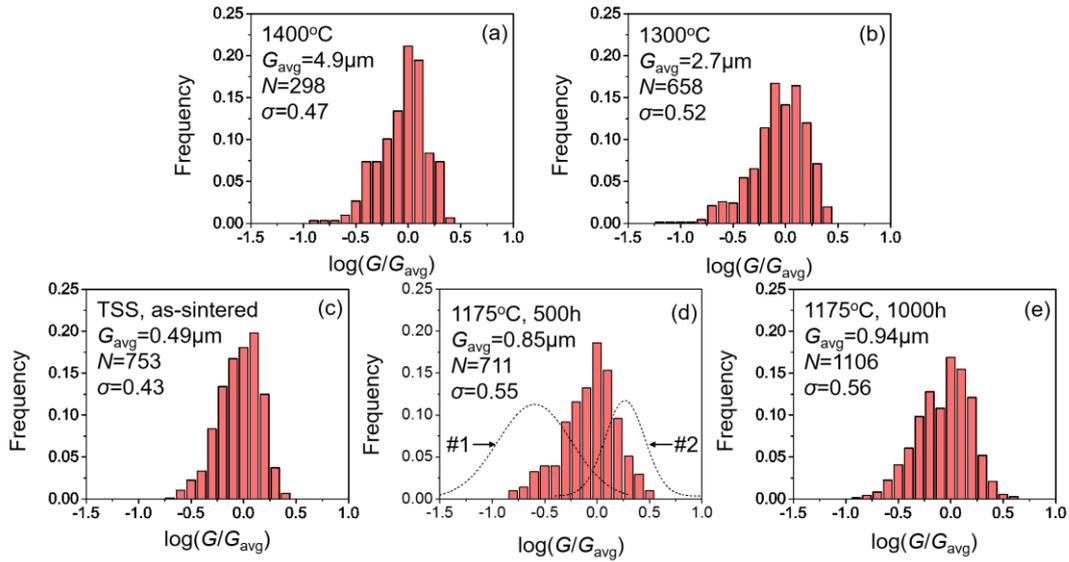

**Figure 9** Normalized grain size distributions of NS samples annealed for 8 h, at (a) 1400 °C, and (b) 1300 °C. Same for TSS samples (c) without post-sintering annealing, (d) annealed at 1175 °C for 500 h and (e) at 1175 °C for 1000 h. Dotted distributions #1 and # in (c) taken from circled regions in **Fig. 8b-c**; their peak heights are arbitrarily chosen for guidance of eye. (These grains not included in background distribution shown as histogram.) Also listed are $G_{avg}$, number $N$ of grains in distribution, and standard deviation $\sigma$ of normalized grain size $G/G_{avg}$.

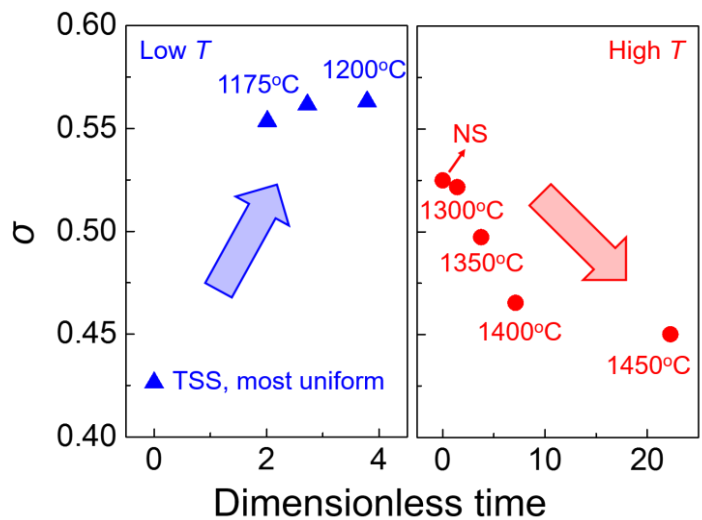

**Figure 10** Bifurcated evolution of standard deviation $\sigma$ of normalized grain size distribution

during post-sintering annealing. Dimensionless time: $\tilde{t} = 2M_b \gamma t = \left(G_{avg}/G_0\right)^2 - 1$. Annealing temperature labeled next to symbol.

## IV. Discussion

**(1) Densification-coarsening competition**

Our data summarized in **Fig. 5** confirmed the mobility transition hypothesized in **Fig. 1**, which was the basis of two-step sintering. Importantly, the transition occurred without a corresponding slowdown in the densification time shown in the Arrhenius plot in the inset of **Fig. 5**. Therefore, without suppressing diffusion, the competition between coarsening and densification is tilted to favor densification as the temperature lowers, which results in a finer grain size in the as-sintered dense ceramics. Two-step sintering takes full advantage of this feature, as evidenced in this study where the most uniform dense microstructure was produced by two-step sintering at 1200 ºC. This is likely because of the lack of grain growth during second-step sintering, so the microstructure can better preserve the feature of the green body. It is an important practical advantage of two-step sintering, in that it can best benefit from the advances in nano-powder technology and quality. This is not always: It is well-known that during conventional sintering, most nano-powders will experience multifold coarsening by the time full densification is reached.

Despite the above advantage, our study made it clear that prolonged post-sintering annealing at either the second-step sintering temperature or a slightly lower temperature could be detrimental to microstructure uniformity. Although the average grain sizes are still very small because of extremely slow grain growth, growing by only ~0.5 μm from a starting size of ~0.5 μm after 500-1000 h, a few clusters of grains of both vanishing grain sizes and very large sizes do appear. These regions could have anomalous properties, thus adversely affecting the reliability of the ceramics. Since as-sintered microstructure is not an equilibrated microstructure in the presence of capillary driving force, prolonged annealing after achieving full densification is not advised.

**(2) Mobility transition**

Regardless of mobility quenching at lower temperature, at 1300 °C and above parabolic grain growth does proceed. This holds for both NS or TSS samples, and during the growth the grain size distribution does evolve to become more "normal", with a gradually decreasing standard deviation (**Fig. 10**). Importantly, the grain size can increase many folds at higher temperatures; in our experiments, NS sample from 1.7 μm to 8.2 μm after 8 h at 1450 °C, and TSS sample from 0.49 μm to 5.7 μm after 64 h at 1300 °C. Correspondingly, the grain boundary area shrinks by 5-12 times, the 3-grain lines by 25-140 times, and the grain number of 4-grain junctions by 125-1700 times. Therefore, if compositional changes of the interfaces/junctions as a result of solute/impurity/second-phase segregation can cause any mobility slowdown, then they should have become very pronounced in these higher temperature experiments. Yet parabolic law is still obeyed, which implies that such changes are either insignificant or inconsequential in our samples. This is not unreasonable because 8YSZ as a cubic zirconia solid solution is known to be very stable, with relatively weak solute segregation, and has a very large solubility for many cations (even for anions, e.g., N). It follows that the observed mobility quenching, grain growth stagnation and the gradual development of inhomogeneous microstructure, including clusters of abnormally small/large grains seen in the lower temperature experiments, are unlikely to be caused by the compositional changes. Instead, they must have a structural origin.

In normal coarsening—taken in the broadest sense not limited to the parabolic growth—the size distribution tends to narrow as the growth exponent $n$ increases[22]. This is best known in the Oswald ripening, which follows the cubic growth law: If nucleation is limited to a one-time event, then the size distribution will asymptotically approach a monodisperse one, which has been seen in the preparation of monosized ceramic powders and nanoparticles. This is because a larger $n$ implies a relatively slower growth rate for the larger grains, and a relatively faster growth rate for the smaller ones, which allows larger ones to slow down and slower ones to catch up. Meanwhile, much smaller grains will rapidly shrink out of existence. In this way, a more narrowly distributed size distribution will result. This argument stands regardless of the origin of the higher $n$: It could come from different diffusion paths for coarsening such as bulk diffusion ($n=3$) in lieu of grain boundary diffusion ($n=2$), or from segregation or pinning of the interface causing a mobility slowdown with size. (We already discarded the latter possibility above.)

Theoretical predictions of grain growth models summarized in **Table 2** confirm the trend[22]: An increasing $n$ is correlated to a smaller ratio of $G_{max}/G_{avg}$ and a decreasing standard deviation $\sigma$ of the normal size. These predictions are for the steady state. If the initial size distribution differs from the steady state one, then it will self-correct and evolve toward the steady state as growth proceeds, whose progress may be measured by the dimensionless time $\tilde{t} = 2M_b\gamma t$, equal to $\left(G_{avg}/G_0\right)^2 - 1$, which is experimentally measurable. This is confirmed with the data at above 1300 °C. Specifically, after 8 h at 1450 °C, we obtained $G_{max}/G_{avg}$ =2.5 vs. the predicted 2.25, and $\sigma$=0.45 vs. the predicted 0.354. This is considered good agreement because the theoretical predictions are for a three-dimensional system whereas our statistics were collected on two-dimensional sections. According to Hillert (his **Fig. 3**[18]) the grain size distribution in two-dimensional sections has a broader tail at large sizes, giving a 10% larger $G_{max}/G_{avg}$.

**Table 2** Model predictions of growth exponent $n$, standard deviation $\sigma$ of $G/G_{avg}$ and $G_{max}/G_{avg}$. Also listed (Experiments) are values found after long annealing.

|  | Oswald ripening | 2-grain boundary control | 3-grain line control | 4-grain junction control | Experiments >1300 °C | <1300 °C |
|---|---|---|---|---|---|---|
| $n$ | 3 | 2 | 1 | exponential | 2 | 5-6 |
| $\sigma$ | 0.215 | 0.354 | 1.0 | No steady-state | 0.45-0.52 | ~0.55 |
| $G_{max}/G_{avg}$ | 1.5 | 2.25 | ∞ | ∞ | 2.3-3.0 | 5-6 |

As the growth temperature drops below 1300 °C, we found growth stagnation and $n$ increasing in **Fig. 7**. However, contrary to the trend seen at higher temperature and in theoretical prediction, both $G_{max}/G_{avg}$ and $\sigma$ actually increase in **Fig. 10**. Even more remarkable is the observation of isolated pockets of abnormally small grains and abnormally large grains in **Fig. 8,** which lie completely outside the norm in **Fig. 9d**. These anomalies cannot be resolved by allowing mixed control, Eq. (4), since we have performed numerical simulations for such model, and in every essential detail of the simulated results (data not shown) we found them contradictory to our experimental observations. These include: (a) The solution shows that after an initial delay, parabolic grain growth resumes because the concentrations of 3-grain lines and 4-grain junctions decrease as the grain size increases. This is contrary to our observation of initially parabolic growth followed by stagnant growth, in **Fig. 7**. (b) The solution shows that the standard deviation initially increases reflecting the trend for higher $\sigma$ in 3-grain-line-controlled growth and 4-grain-junction-controlled growth shown in **Table 2**, but it later decreases as the concentration of these lines and junctions diminishes over time. This is contrary to our observation of finding abnormal clusters only after 500 h of slow grain growth. Therefore, generalizing the mean field theory to allow multiple mobility limitations cannot explain our data.

**(3) Mobility inhomogeneities and microstructure dispersion**

Although the simple theories described above in Eq. (1-4) fail to explain our low temperature data, we now show that a new model that allows inhomogeneities in grain boundary properties can. In this model, we allow a sub-population of grains to exist that have immobile 2-grain boundary, 3-grain lines or 4-grain junctions. As a result, such grains are unable to grow and will stay dormant, as if they are "by-standers", while "watching" other grains in the other sub-population grow. In the simplest model of this kind, the two sub-populations do not interact at all, so mass conservation within each sub-population holds and their grain growths (the dormant sub-population having none) are completely independent of each other. Obviously, the grain size dispersion of the entire population will worsen with time. Moreover, as the average grain size is weighed down by the dormant sub-population, whose number of grains remains unchanged, the growth will appear to experience stagnation, the more so the larger (hence fewer, which is the case at longer time) the average grains of the growing sub-population. These qualitative predictions are in agreement with the broad statistical trend observed in **Fig. 7** and **Fig. 10**. Assuming bimodal mobilities, we have verified the above mechanism by numerical simulations shown in **Fig. 11** (stagnation is clear from the $G_{avg}^2$-$t$ plot in **Fig. 11a**, while the microstructure becomes more inhomogeneous with an increasing $\sigma$ in **Fig. 11b**; see Ref. 22 for the details), and found that even less extreme assumptions in the model will still lead to qualitatively similar features (data not shown; see Ref. 22 for more details.) The origin of the immobile subpopulation is likely to be structural as we have already argued, but phenomenologically it can be described by **Fig. 1** by assigning a larger activation energy to the mobilities of grains in such sub-population. As such, they become increasingly important at lower temperature, which is consistent with our experiments.

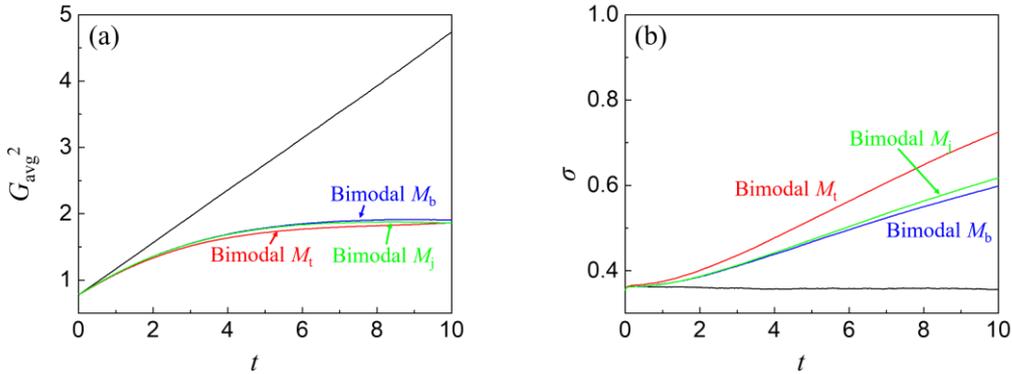

**Figure 11** Calculated (a) $G_{avg}^2$ and (b) $\sigma$ as a function of time $t$, with reference curves (in black) following parabolic growth law for unimodal mobility ($M_b=M_t=M_j=M$), and color curves for bimodal mobilities: for mobile boundaries/junctions, $M_b=M_t=M_j=M$, and for immobile ones, either $M_b=M/10^4$ (in blue), $M_t=M/10^4$ (in red), or $M_j=M/10^8$ (in green).

To further distinguish which structure—2-grain boundary, 3-grain line or 4-grain junction—is most likely to explain our data, we will return to two additional key observations of post-TSS annealing: (a) An incubation time (~8 h) before grain growth ensues was observed in $Y_2O_3$ of 100 nm grain size[2] but not in our study of 0.49 μm YSZ; and (b) clusters of abnormally small grains form after prolonged annealing. These observations are consistent with a dominant role of 4-grain junctions for the following reasons. First, when 4-grain junctions control the growth, there is an incubation time before exponential growth, which was observed in (a). The reason such incubation time was not seen in our grain-size data in **Fig. 7** may be explained by the larger grain size, 0.49 μm in YSZ compared to 100 nm in $Y_2O_3$, which amounts to 125 times different concentrations of 4-grain junctions, making them much less important in YSZ. Second, while many models[25-29] can explain why clusters of large grains form, only 4-grain-junction-controlled growth can allow small grain clusters to form because small grains cannot disappear in this model, which gives $dG/dt$=0 at $G$=0 (see Eq. (3)). Therefore, as large grains grow away to leave sluggish 4-grain junctions behind, a cluster of small dormant grains will remain.

Further insight into the kinetics may be gleaned from our observation that the large/small grain clusters do not form until after 500 h annealing at 1200°C or 1175°C. Taking 500 h as the transient time $\tau$ before inhomogeneous mobilities can significantly affect the local microstructure, we estimate the inhomogeneities $M_b^*$, $M_t^*$, and $M_j^*$, from the transient time in the **Introduction**: $\tau \sim G^2/M_b\gamma$ for the 2-grain boundary mobility, $\tau \sim Ga/M_t\gamma$ for the 3-grain line mobility, and $\tau \sim a^2/M_j\gamma$ for the 4-grain junction mobility. In the above, $G$ is taken as the prevailing grain size, which is 0.49 μm in **Fig. 8**. Taking $a$=0.5 nm and $\gamma$ =0.3 J/m², we estimate these inhomogeneities to be $M_b^*$=4.4×10$^{-19}$, $M_t^*$=4.5×10$^{-22}$, and $M_j^*$=4.6×10$^{-25}$, all in unit of m⁴·s⁻¹·J⁻¹. ($M_b^*$ can also be read from **Fig. 5** below the temperature of mobility transition, which gives comparable values, but these estimates for $M_t^*$ and $M_j^*$ are new.) As already argued above, the appearance of small grain cluster favors the interpretation of $M_j^*$ that characterizes a subpopulation of extremely immobile 4-grain junctions, and we suspect it is the transient $\tau \sim a^2/M_j\gamma$ that determines the appearance of small/large grain clusters. Physically, whereas most regions are likely to consist of some isolated immobile junctions thus sensing an intermediate pinning effect, we may envision some extreme though rare local regions containing mostly immobile junctions to evolve into microstructures shown in **Fig. 8b**.

## V.    Conclusions

(1) A finer grain size in as-sintered dense samples is correlated with a lower sintering temperature, because the competition between coarsening and densification is tilted to favor densification as the temperature lowers.

(2) The apparent grain boundary mobility decreases rapidly below 1300 °C with an apparent activation energy of ~10 eV, more than twice the value of 4.2 eV above 1300 °C.

(3) The grain-size distribution is the narrowest in the two-step sintered sample in the as-sintered state, and it broadens with low-temperature annealing despite little average grain growth is produced; it even features local pockets of either abnormally large or abnormally small grains. Such size-distribution broadening, size heterogeneities and growth stagnation do not happen at higher temperature.

(4) These results are consistent with a heterogeneous distribution of grain boundary/grain junction mobilities, which becomes more heterogeneous at lower temperatures because of different activation energies. Unlike any other grain-growth theory, such model can uniquely predict growth stagnation and increased grain size dispersion.

**Acknowledgements**

This work was supported by the Department of Energy (BES grant no. DEFG02-11ER46814) and used the facilities (LRSM) supported by the U.S. National Science Foundation (grant no. DMR-1120901).

interface structure. *J. Am. Ceram. Soc.* 2009;92;1464-1471.